# Shallow NV centers augmented by exploiting n-type diamond

A. Watanabe[1], T. Nishikawa[1], H. Kato[2], M. Fujie,[1] M. Fujiwara[1], T. Makino[2], S. Yamasaki[2], E. D. Herbschleb[1], N. Mizuochi[1*]

[1]Institute for Chemical Research, Kyoto University, Gokasho, Uji-city, Kyoto 611-0011, Japan

[2]National Institute of Advanced Industrial Science and Technology (AIST), Tsukuba, Ibaraki 305-8568, Japan

*Corresponding author. Email: mizuochi@scl.kyoto-u.ac.jp

**Abstract**

Creation of nitrogen-vacancy (NV) centers at the nanoscale surface region in diamond, while retaining their excellent spin and optical properties, is essential for applications in quantum technology. Here, we demonstrate the extension of the spin-coherence time ($T_2$), the stabilization of the charge state, and an improvement of the creation yield of NV centers formed by the ion-implantation technique at a depth of ~15 nm in phosphorus-doped n-type diamond. The longest $T_2$ of about 580 μs of a shallow NV center approaches the one in bulk diamond limited by the nuclear spins of natural abundant $^{13}$C. The averaged $T_2$ in n-type diamond is over 1.7 times longer than that in pure non-doped diamond. Moreover, the stabilization of the charge state and the more than twofold improvement of the creation yield are confirmed. The enhancements for the shallow NV centers in an n-type diamond-semiconductor are significant for future integrated quantum devices.

# 1. Introduction

A negatively charged NV center in diamond is a versatile atomic-sized spin system for remarkable applications in quantum sensing[1-8] and quantum-information science.[9-13] The reason is that it has excellent properties such as long coherence times[8] and high sensitivities with nanometer-scale resolution. By utilizing nanoscale shallow NV centers, applications for nanoscale imaging[5,6] and nanoscale nuclear magnetic resonance[1-4] were demonstrated. Furthermore, NV centers were successfully integrated into photonic and mechanical structures,[7,11,14-16] and in electronic devices,[12,13,17-21] which expands their versatility.

In these applications, the NV centers are required to be produced with nanometer spatial accuracy, while retaining their excellent spin and optical properties. Among the techniques to generate NV centers, the ion-implantation technique is the most effective for producing them very close to the surface, and for fabrication of arrays with nanoscale resolution.[22,23] However, this technique has drawbacks when comparing the ion-implanted NV centers at nanoscale depths with those in bulk diamond. Namely, they have much shorter coherence times, and their charge states become unstable due to defects created during the ion implantation in the surface region.[24-26]

So far, extremely shallow NV centers, which have depths less than 5 nm, have been investigated significantly, because they are required for many applications. However, they are strongly affected by surface defects and surface states. [27-30] On the other hand, for NMR applications [1], slightly deeper NV centers, at a depth of 25 nm, are utilized to suppress the NMR signal broadening caused by spin diffusions.[30-32]

Recently, we investigated NV centers in phosphorus-doped n-type diamond which were created during growth by the chemical-vapor deposition (CVD) technique. We looked at the stabilization of the charge state[33,34] and the extension of the electron spin-coherence times ($T_2$, $T_2^*$).[8] With respect to stabilizing the negatively charged state of the NV center (NV$^-$),[33,34] the stochastic charge-state transitions between NV$^-$ and the neutral charge state (NV$^0$), visible as blinking of the fluorescence of NV centers during weak laser irradiation,[33-35] were suppressed. With respect to the coherence times, the longest $T_2$ and $T_2^*$ ($T_2 \cong 2.4$ ms, $T_2^* = 1.5$ ms) among electron spins in room-temperature solid-state systems were demonstrated.[8] This is against intuition, because phosphorus is paramagnetic at room temperature and thus causes magnetic noise. The extensions of $T_2$ and $T_2^*$ were attributed to the suppression of the creation of multi-vacancy complex or vacancy-impurity complex defects during CVD growth by Coulomb repulsion among charged defects.[8]

Regarding the ion implantation method, improvements of $T_2$ and the creation yield of shallow NV centers were reported.[36,37] In one research,[36] a thin boron-doped p-type diamond was deposited as a sacrificial film on non-doped diamond to suppress defect formation during nitrogen ion-implantation. After ion-implantation, annealing and subsequent removal of the p-type diamond, shallow NV centers remained at the non-doped diamond's surface. In other research,[37] nitrogen was ion-implanted into diamond where impurities such as sulfur, oxygen, or phosphorus were ion-implanted in advance to suppress the formation of complexes during the later implantation by charging defects. From the viewpoint of suppression by Coulomb repulsion among the charged defects, the higher quality of n-type diamond is expected to be more effective. So far, n-type high Hall mobility in diamond was realized by CVD growth with phosphorus doping.[38,39] In this research, we investigate $T_2$, the charge-state stability, and the creation efficiency of single shallow NV centers formed by ion-implantation into a phosphorus-doped n-type diamond grown by the CVD technique.

## 2. Material and Method

A thin film of phosphorus-doped diamond is grown by microwave plasma-enhanced CVD onto a IIa (111) HPHT (high-pressure and high-temperature, 0.3 mm thickness) diamond substrate for sample I. It is grown to achieve a phosphorus concentration of $5 \times 10^{16}$ cm$^{-3}$ and a thickness of about $7 \times 10^2$ nm using methane with a natural abundance of $^{13}$C (1.1%). The growth condition is almost the same as in previous research except for the $^{13}$C concentration.[8,39] The details are shown in the supporting information.[40]

The NV centers are formed at the surface region of the n-type diamond. To create these shallow NV centers, nitrogen isotope ($^{15}$N) ions are simultaneously implanted to both sample I and sample II, the latter being a pure IIa (111) HPHT diamond as reference (see Figure 1(a, b)). The implanted density is around $5 \times 10^8$ atoms cm$^{-2}$ with a kinetic energy of 10 keV at 600 °C. After annealing at 800 °C for 30 minutes to produce the NV centers, the samples are washed by hot acid.[40] This annealing temperature is lower than that in previous reports[41] to suppress potential change in electrical character of the sample.

The depths of the $^{15}$N ions are simulated by an ion-implantation Monte-Carlo simulator (SRIM) as shown in Figure 2(a). Its peak is at a depth of around 15 nm. The NV centers are measured by an in-house built confocal microscope, Figure 1(c, d) display example images. The bright spots are single NV centers as confirmed by anti-bunching measurements, an example is given in Figure 2(b). In the confocal microscope, two acoustic optical modulators were utilized to suppress the background illumination during the echo measurements.

With the optically detected magnetic resonance (ODMR) spectra of single NV centers, we can confirm whether they are produced by the ion-implantation. In case of the implanted $^{15}$N isotope, the ODMR spectrum shows hyperfine splitting with two lines about 3 MHz apart due to its nuclear quantum number of 1/2[42] as shown in Figure 2(c). On the other hand, in case of $^{14}$N, which natural abundance is about 99.6 %, the ODMR spectrum shows hyperfine splitting with three lines each about 2 MHz apart due to its nuclear quantum number of 1 as shown in Figure 2(d). In Figure 1 (c,d), both $^{15}$NV and $^{14}$NV are visible. In sample I, most of the observed NV centers had the $^{15}$N isotope, while in sample II, more than half of the NV centers had the $^{14}$N isotope. Throughout the presented research, the implanted NV centers are investigated.

For electron paramagnetic resonance (EPR) spectra of a substitutional nitrogen (P1 center),[43] we use a Bruker ELEXSYS X-band spectrometer. The spin concentration is estimated by comparing the integral of the signal intensity with that of a Bruker strong pitch standard sample.

## 3. Results and Discussion

The coherence times ($T_2$) of single NV centers in samples I and II are measured by Hahn-echo sequences as shown in Figure 3(a). In order to remove common-mode noise, the phase of the last π/2-pulse is inverted as depicted by the ±-sign in Figure 3(a).[8] The applied static magnetic field is 3.2 mT in most experiments. Revivals of the echo decays due to the $^{13}$C nuclear spin precession are observed in the echo signals, which rate matches the Larmor precession frequency for the $^{13}$C nuclear spin of 10.705 MHz T$^{-1}$.[44] The results of the decay are fitted to the exponential $\exp(-(\tau/T_2)^n)$.[8]

The Hahn-echo decays for the longest $T_2$ and an average $T_2$ of single NV centers in each sample are shown in Figure 3(b, c). The longest $T_2$ and the averaged $T_2$ in sample I are 579.0 ± 28.7 μs and 325.7 ± 148.2 μs respectively, as summarized in Table 1. The longest $T_2$ and the averaged $T_2$ in sample II are 359.6 ± 10.3 μs and 184.6 ± 76.5 μs respectively. It is indicated that both the longest and averaged $T_2$s in sample I are longer than those in sample II. As far as we know, these are longer than the reported ones of the nanoscale shallow NV centers created by ion-implantation with similar depth.[24,25,36,37,45,46] However, $T_2$ severely depends on the depth and the impurity concentration, so it is hard to warrant a fair comparison, hence we look at the results in more detail below.

The longest $T_2$ in sample I is comparable with the one in bulk diamond that is limited by the $^{13}$C nuclear spin bath (∼0.6 ms) given a natural abundance concentration.[47,48] On the other hand, the averaged $T_2$ is shorter than it. If there are no noise sources other than the $^{13}$C nuclear spin in sample I, $T_2$ should be about 0.6 ms, because the $^{13}$C concentration in sample I follows from natural abundance. Since, apart from this concentration, all other growth conditions of sample I are almost the same as those in our previous research[8,40], the main noise sources besides the $^{13}$C nuclear spins are considered to be paramagnetic defects generated by the ion-implantation such as multi-vacancy complex or vacancy-impurity complex defects or surface defects.[35] Besides, it is observed that $T_2$ is relatively scattered, as seen from the large uncertainties of the averaged $T_2$. This may be attributed to different environmental noises from bulk defects and/or surface defects near the NV centers. Particularly, the scatter of sample I is larger than that of sample II, which may be attributed to the additional environmental noise from phosphorus atoms, that are accidentally close to a specific NV center.

Next, we discuss the potential noise sources in sample II. Similar to sample I, one candidate for the main noise source in sample II is considered to be the paramagnetic defect generated during the ion-implantation. Otherwise, the possibility of a substitutional nitrogen (P1 center) can be considered as a candidate, because it can easily incorporate into diamond[48] and its neutral charge state is the electron doublet state ($S=1/2$). Therefore, we investigate the concentration of the P1 centers in sample II by EPR.[43] In Figure 4(a), the EPR spectrum for sample II is shown. The concentration of the P1 centers in sample II is estimated to be $(2.6 \pm 0.1) \times 10^{16}$ cm$^{-3}$ (~0.14 ppm).

In addition to nitrogen, boron is also a well-known impurity that can easily incorporate into diamond. However, from the observation of the EPR signal of the P1 center, we can exclude the possibility of boron as a main noise source. If boron simultaneously exists with the P1 center, the charge state of the P1 center changes from neutral to positive while the charge state of boron changes from neutral to negative, which makes both non-paramagnetic centers. Therefore, the observation of the EPR signal of the P1 center indicates that the number of P1 centers is larger than that of boron.

In previous research about $T_2$ of the NV center,[48] the dependency of $T_2$ on the P1 concentration was investigated. From their results, in case of a P1 concentration of $2.6 \times 10^{16}$ cm$^{-3}$, $T_2$ is expected to be longer than 400 μs, which is much longer than the averaged $T_2$ of 184.6 μs in sample II. In their report, below a P1 concentration of 1 ppm, it was shown that $T_2$ is inversely proportional to the P1 concentration, and a relationship for $T_2$ of 160±12 μs·ppm was estimated. From this relationship, if the averaged $T_2$ of 184.6 μs in sample II is limited by P1, the concentration of P1 centers should be 0.87 ppm, which is about six times larger than the P1 concentration of ~0.14 ppm estimated for sample II. From these results, the main noise sources in sample II are considered to be paramagnetic defects generated by the ion-implantation, just like the main noise sources in sample I. Therefore, the difference in $T_2$ between samples I and II is considered to originate from the effect of the n-type diamond on the defect-generation during ion-implantation. Namely, similar to previous researches,[8,36,37] our results suggest that the generation of multi-vacancy complex or vacancy-impurity complex defects is suppressed by Coulomb repulsion of charged vacancies or impurities in n-type diamond.

In previous related reports, twofold[36] and tenfold[37] improvements of the NV creation yield by nitrogen ion-implantation were reported. We compare the yield studying the NV density by counting the NV centers created by $^{15}$N ion-implantation in a unit area. The $^{15}$NV densities are shown in Table 1. In our results, the NV density of sample I is more than two times larger than that of sample II, thus demonstrating the effectiveness of defect-charging on the NV creation yield during the nitrogen ion-implantation. The smaller yield compared with the tenfold improvement[37] might be due to a dependence of the effects on the depth, since their NV centers are 4 times deeper than ours.

We investigate the stabilization of the charge state of $NV^-$ by single-shot charge-state measurements. In previous research about the stabilization of the charge state of $NV^-$ in phosphorus-doped n-type diamond grown by the CVD technique,[33,34] the stabilization effect was shown with single-shot charge-state measurements.[33-35]

We perform nondestructive single-shot charge-state measurements to compare the stability of the charge state of $NV^-$ in both samples. The details of the measurement are summarized in the supporting information.[40] Figure 5(a) shows a typical time trace of the fluorescence from a single NV center in sample I under weak laser irradiation (3-μW, 593-nm laser), which shows some sudden jumps due to the interconversions between $NV^0$ and $NV^-$. Figure 5(b) shows the single-shot charge-state measurement sequence. The histograms of the charge-state population of a single NV center are plotted in Figures

5(c) and 5(d). The histograms are fitted by a double Poisson distribution and the populations of $NV^0$ and $NV^-$ are estimated from the area of the two peaks. The averages and standard deviations of the $NV^-$ population in samples I and II are extracted by fitting Gaussian function to be $0.746 \pm 0.017$ and $0.747 \pm 0.019$ respectively, as shown in Table I. In the case of non-doped diamond, the steady-state $NV^-$ population is always less than about 80% under 532 nm initialization.[33-35] On the other hand, we find 8 NV centers with an $NV^-$ population of over 0.8 in sample I among 78 measured NV centers; Figure 5(c) shows the histogram with the highest $NV^-$ population (0.92). The almost 10% of NV centers with an $NV^-$ population of over 0.8 is significant, as this is over 3 standard deviations away from the average, which should represent only 0.1% in a normal distribution. This result supports that the phosphorus-doped *n*-type diamond stabilizes $NV^-$ in sample I. In the case of the NV centers produced during growth in phosphorus-doped n-type diamond with the CVD technique, it is reported that the $NV^-$ population becomes almost 100%.[33] The smaller $NV^-$ population in the present research is probably due to the effect of damage caused by the nitrogen ion-implantation and/or surface states[49]. Regarding the surface states, they can compensate the phosphorus donor electrons, which can lower the charge stability at this depth.

Compared with the previous results with respect to the extension of $T_2$ and the charge state stabilization in phosphorus doped n-type diamond grown by the CVD technique,[8, 33] here, n-type doping seems less effective in the near-surface regime. However, we believe that they can be improved by, for example: optimization of the surface termination, the phosphorus concentration, or the dopant profile by using delta-doping or a super-lattice.

## 4. Conclusion

In conclusion, we investigated NV centers created by the ion-implantation technique at ~15 nm depth from the surface in phosphorus-doped n-type diamond, which was grown with the CVD technique. We demonstrated the extension of $T_2$, the stabilization of the charge state, and the improvement of the creation yield of the NV centers compared with those generated by the same ion-implantation technique in non-doped diamond. As far as we know, the enhanced $T_2$ is longer than in other reported nanoscale shallow NV centers created by ion-implantation at similar depth. Additional significance stems from the usage of a diamond semiconductor to obtain the enhancements. The present result is important for nano-fabrication of integrated devices, so it paves the way to the development and application of diamond-based quantum-information, sensing, and spintronic devices.

**Acknowledgements**

The authors acknowledge the financial support from KAKENHI (16H06326), MEXT Q-LEAP (No. JPMXS0118067395), and the Collaborative Research Program of the Institute for Chemical Research, Kyoto University (2020-110).

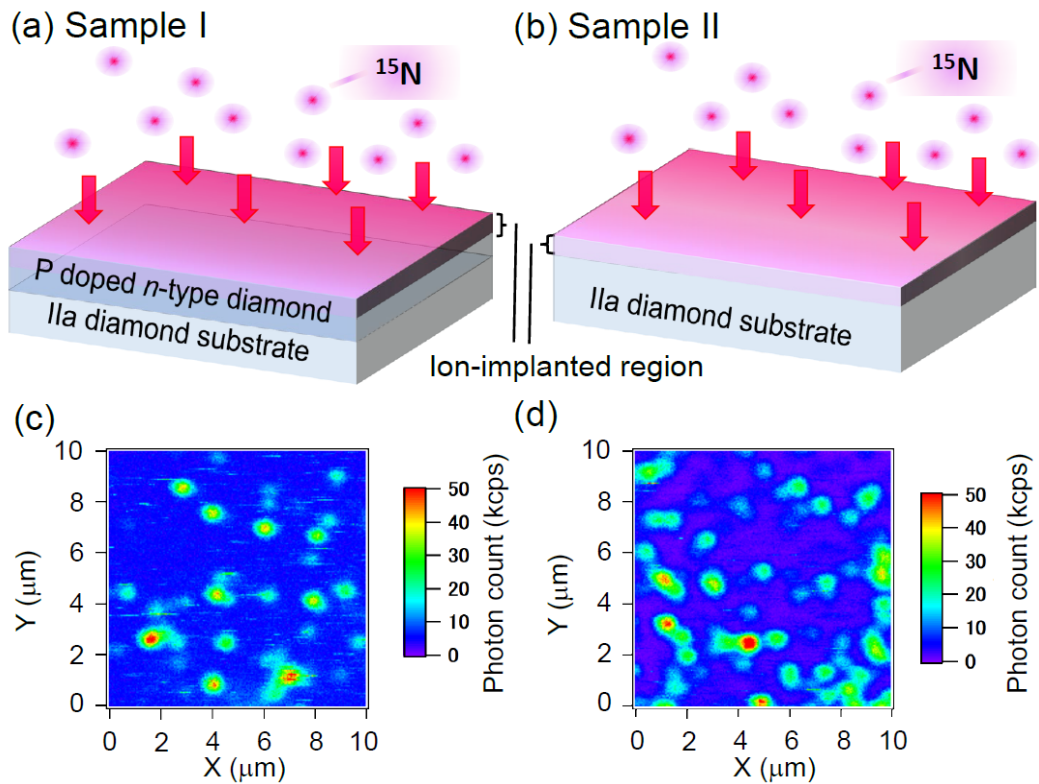

Figure 1. Schematic images of samples I (a) and II (b). In sample I, phosphorus-doped n-type diamond with a thickness of about $7 \times 10^2$ nm is deposited. The peak of the concentration of the NV centers is at a depth of around 15 nm. Confocal microscope scan images of samples I (c) and II (d).

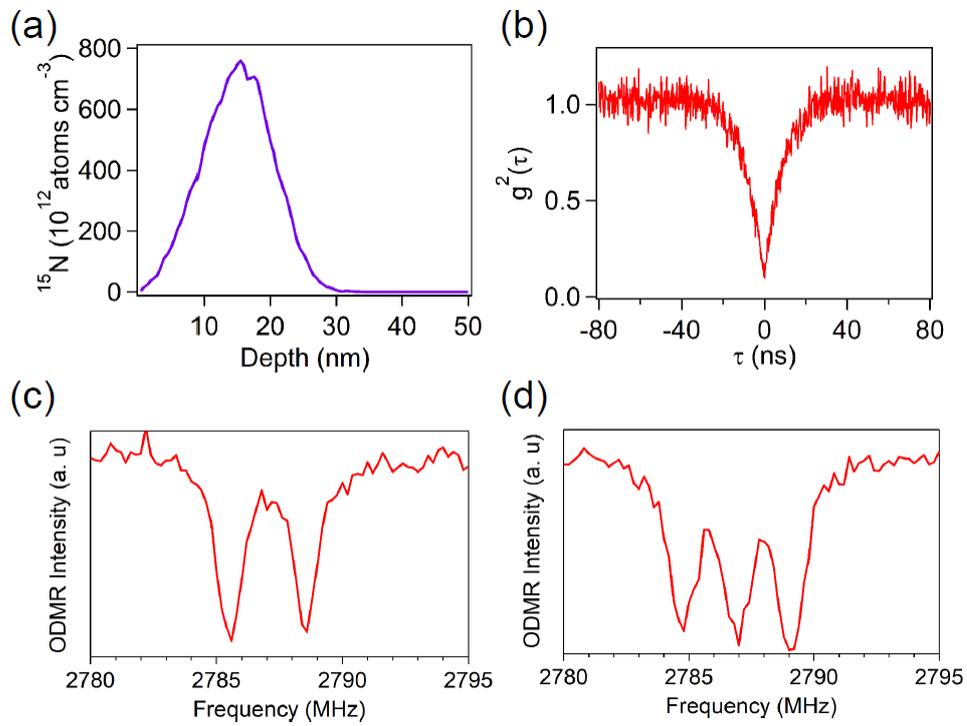

Figure 2. (a) Simulation result of an ion-implantation Monte-Carlo simulator (SRIM). (b) Result of an anti-bunching measurement. ODMR spectra for NV centers with $^{15}$N (c) and $^{14}$N (d) isotopes.

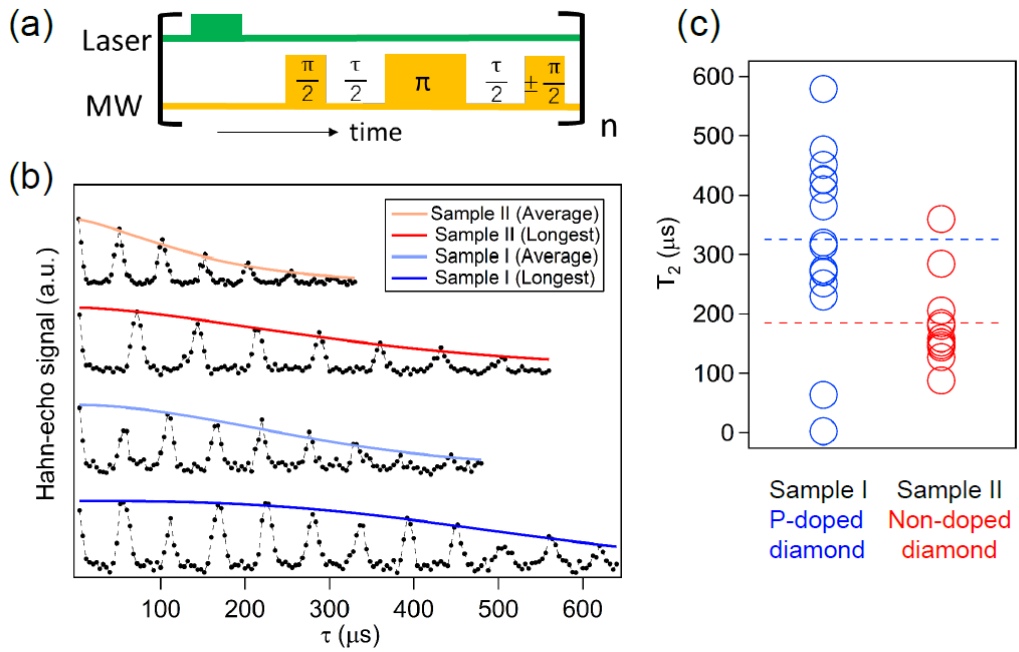

Figure 3. (a) Pulse sequence for a Hahn-echo measurement. (b) Hahn-echo decays of NV centers in samples I and II. Echo decays of the longest and an average length $T_2$ in each sample are shown. The revivals in echo decays are due to $^{13}$C nuclear spin precession, which rate matches the Larmor precession frequency for the $^{13}$C nuclear spin. (c) The $T_2$ of samples I (left blue) and II (right red). The average length of $T_2$ in each sample is indicated with a dotted line.

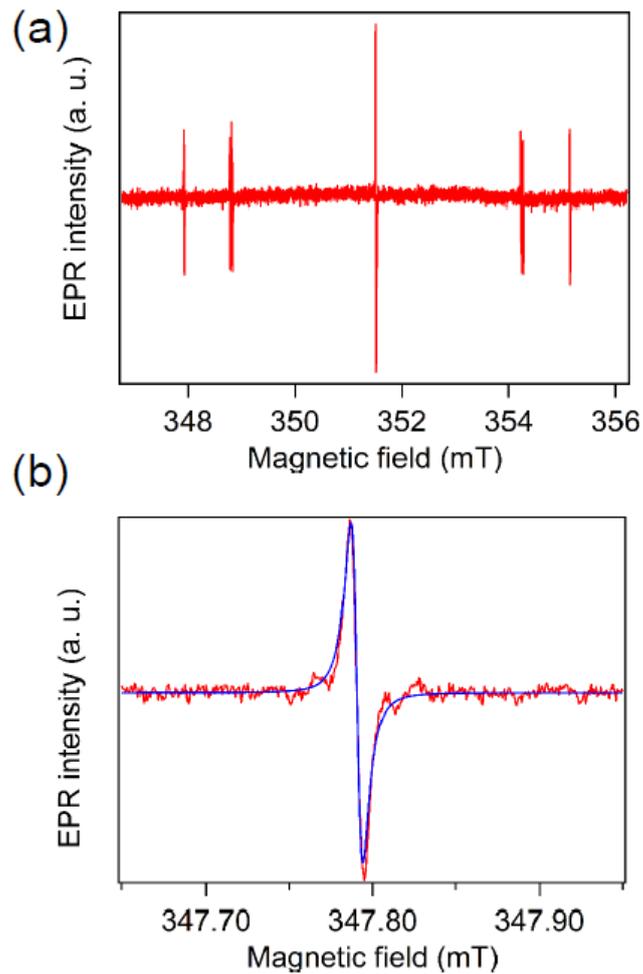

Figure 4. (a) The EPR spectrum for sample II to study the P1 center, with (b) showing an expanded region. The microwave frequency is 9.8546 GHz. The microwave power and the modulation amplitude are 0.2 µW and 4 µT, respectively. The direction of the magnetic field is almost parallel to the [111] axis. The blue line follows from the derivative of a fitted Gaussian.

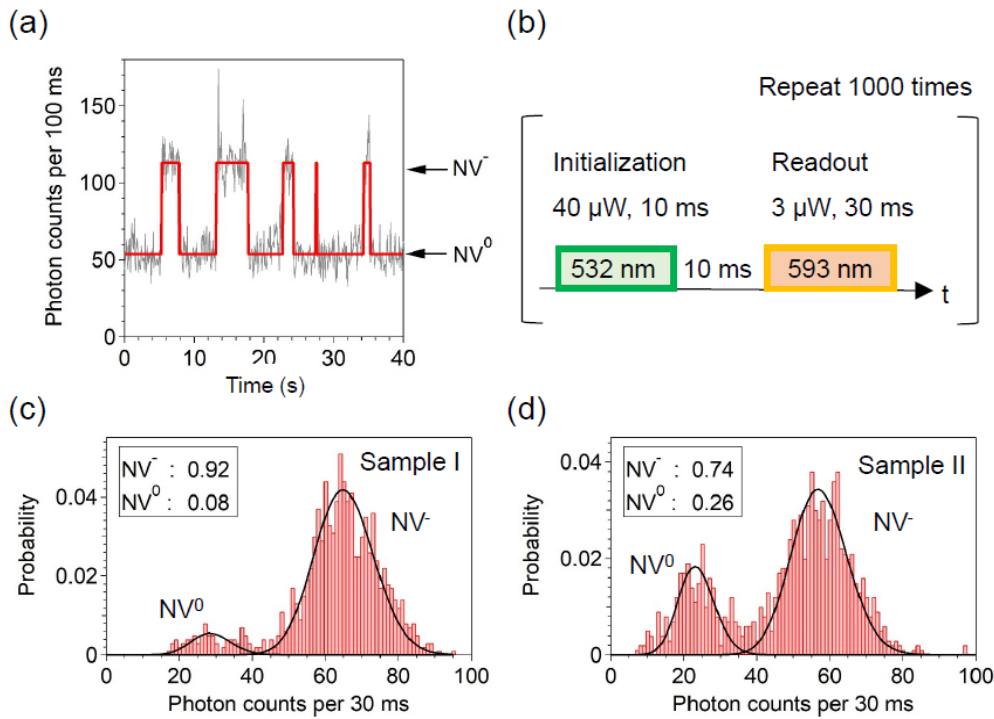

Figure 5. (a) Typical time trace of the fluorescence from a single NV center in Sample I under continuous 3-µW, 593-nm laser illumination. Sudden jumps in the photon count correspond to the interconversions between $NV^0$ and $NV^-$. The solid red line shows the most probable fluorescence levels of $NV^0$ and $NV^-$ as obtained by a hidden Markov model. (b) Measurement sequence of the nondestructive single-shot charge-state measurements. Using this sequence for a single NV center, the histogram displaying the $NV^0$ and $NV^-$ populations is obtained. (c) and (d) Histograms for the NV centers with the highest $NV^-$ populations measured in samples I and II.

Table 1

|  | Phosphorus-doped n-type diamond (Sample I) | Non-doped diamond (Sample II) |
|---|---|---|
| Average $T_2$ | 325.7 ± 148.2 µs | 184.6 ± 76.5 µs |
| Longest $T_2$ | 579.0 ± 28.7 µs | 359.6 ± 10.3 µs |
| $^{15}$NV density | (27.5 ± 17.1) / 100 µm² | (12.5 ± 9.6) / 100 µm² |
| Average population ratio of NV⁻/NV⁰ | 0.746 ± 0.017 | 0.747 ± 0.019 |
| Numbers of NV⁻ population of over 0.8 / measured NV | 8 / 78 | 0 / 56 |